\documentclass[%
preprint,
superscriptaddress,
nofootinbib,
 amsmath,amssymb,
 aps]{revtex4-2}
\usepackage{graphicx}                               
\usepackage{textcomp}
\usepackage{xcolor}
\usepackage{dcolumn}                                
\usepackage{csquotes}
\usepackage{bm}                                     
\usepackage[version-1-compatibility, load=prefixed, load=abbr, binary-units=true, separate-uncertainty=true]
{siunitx}                                           
\usepackage{booktabs}                                 
\usepackage{multirow}                                  

\usepackage{comment}
\usepackage{subfig}

\usepackage[mathlines]{lineno}

\usepackage{etoolbox}
\makeatletter
\patchcmd\linenumberpar{\@LN@parpgbrk}{\penalty\@LN@parpgpen\relax}{}{}
\makeatother

\begin{document}



\title{Search for GeV Neutrino Emission During Intense Gamma-Ray Solar Flares with the IceCube Neutrino Observatory}

 \affiliation{III. Physikalisches Institut, RWTH Aachen University, D-52056 Aachen, Germany} \affiliation{Department of Physics, University of Adelaide, Adelaide, 5005, Australia} \affiliation{Dept. of Physics and Astronomy, University of Alaska Anchorage, 3211 Providence Dr., Anchorage, AK 99508, USA} \affiliation{Dept. of Physics, University of Texas at Arlington, 502 Yates St., Science Hall Rm 108, Box 19059, Arlington, TX 76019, USA} \affiliation{CTSPS, Clark-Atlanta University, Atlanta, GA 30314, USA} \affiliation{School of Physics and Center for Relativistic Astrophysics, Georgia Institute of Technology, Atlanta, GA 30332, USA} \affiliation{Dept. of Physics, Southern University, Baton Rouge, LA 70813, USA} \affiliation{Dept. of Physics, University of California, Berkeley, CA 94720, USA} \affiliation{Lawrence Berkeley National Laboratory, Berkeley, CA 94720, USA} \affiliation{Institut f{\"u}r Physik, Humboldt-Universit{\"a}t zu Berlin, D-12489 Berlin, Germany} \affiliation{Fakult{\"a}t f{\"u}r Physik {\&} Astronomie, Ruhr-Universit{\"a}t Bochum, D-44780 Bochum, Germany} \affiliation{Universit{\'e} Libre de Bruxelles, Science Faculty CP230, B-1050 Brussels, Belgium} \affiliation{Vrije Universiteit Brussel (VUB), Dienst ELEM, B-1050 Brussels, Belgium} \affiliation{Department of Physics and Laboratory for Particle Physics and Cosmology, Harvard University, Cambridge, MA 02138, USA} \affiliation{Dept. of Physics, Massachusetts Institute of Technology, Cambridge, MA 02139, USA} \affiliation{Dept. of Physics and Institute for Global Prominent Research, Chiba University, Chiba 263-8522, Japan} \affiliation{Department of Physics, Loyola University Chicago, Chicago, IL 60660, USA} \affiliation{Dept. of Physics and Astronomy, University of Canterbury, Private Bag 4800, Christchurch, New Zealand} \affiliation{Dept. of Physics, University of Maryland, College Park, MD 20742, USA} \affiliation{Dept. of Astronomy, Ohio State University, Columbus, OH 43210, USA} \affiliation{Dept. of Physics and Center for Cosmology and Astro-Particle Physics, Ohio State University, Columbus, OH 43210, USA} \affiliation{Niels Bohr Institute, University of Copenhagen, DK-2100 Copenhagen, Denmark} \affiliation{Dept. of Physics, TU Dortmund University, D-44221 Dortmund, Germany} \affiliation{Dept. of Physics and Astronomy, Michigan State University, East Lansing, MI 48824, USA} \affiliation{Dept. of Physics, University of Alberta, Edmonton, Alberta, Canada T6G 2E1} \affiliation{Erlangen Centre for Astroparticle Physics, Friedrich-Alexander-Universit{\"a}t Erlangen-N{\"u}rnberg, D-91058 Erlangen, Germany} \affiliation{Physik-department, Technische Universit{\"a}t M{\"u}nchen, D-85748 Garching, Germany} \affiliation{D{\'e}partement de physique nucl{\'e}aire et corpusculaire, Universit{\'e} de Gen{\`e}ve, CH-1211 Gen{\`e}ve, Switzerland} \affiliation{Dept. of Physics and Astronomy, University of Gent, B-9000 Gent, Belgium} \affiliation{Dept. of Physics and Astronomy, University of California, Irvine, CA 92697, USA} \affiliation{Karlsruhe Institute of Technology, Institute for Astroparticle Physics, D-76021 Karlsruhe, Germany } \affiliation{Dept. of Physics and Astronomy, University of Kansas, Lawrence, KS 66045, USA} \affiliation{SNOLAB, 1039 Regional Road 24, Creighton Mine 9, Lively, ON, Canada P3Y 1N2} \affiliation{Department of Physics and Astronomy, UCLA, Los Angeles, CA 90095, USA} \affiliation{Department of Physics, Mercer University, Macon, GA 31207-0001, USA} \affiliation{Dept. of Astronomy, University of Wisconsin{\textendash}Madison, Madison, WI 53706, USA} \affiliation{Dept. of Physics and Wisconsin IceCube Particle Astrophysics Center, University of Wisconsin{\textendash}Madison, Madison, WI 53706, USA} \affiliation{Institute of Physics, University of Mainz, Staudinger Weg 7, D-55099 Mainz, Germany} \affiliation{Department of Physics, Marquette University, Milwaukee, WI, 53201, USA} \affiliation{Institut f{\"u}r Kernphysik, Westf{\"a}lische Wilhelms-Universit{\"a}t M{\"u}nster, D-48149 M{\"u}nster, Germany} \affiliation{Bartol Research Institute and Dept. of Physics and Astronomy, University of Delaware, Newark, DE 19716, USA} \affiliation{Dept. of Physics, Yale University, New Haven, CT 06520, USA} \affiliation{Dept. of Physics, University of Oxford, Parks Road, Oxford OX1 3PU, UK} \affiliation{Dept. of Physics, Drexel University, 3141 Chestnut Street, Philadelphia, PA 19104, USA} \affiliation{Physics Department, South Dakota School of Mines and Technology, Rapid City, SD 57701, USA} \affiliation{Dept. of Physics, University of Wisconsin, River Falls, WI 54022, USA} \affiliation{Dept. of Physics and Astronomy, University of Rochester, Rochester, NY 14627, USA} \affiliation{Oskar Klein Centre and Dept. of Physics, Stockholm University, SE-10691 Stockholm, Sweden} \affiliation{Dept. of Physics and Astronomy, Stony Brook University, Stony Brook, NY 11794-3800, USA} \affiliation{Dept. of Physics, Sungkyunkwan University, Suwon 16419, Korea} \affiliation{Institute of Basic Science, Sungkyunkwan University, Suwon 16419, Korea} \affiliation{Dept. of Physics and Astronomy, University of Alabama, Tuscaloosa, AL 35487, USA} \affiliation{Dept. of Astronomy and Astrophysics, Pennsylvania State University, University Park, PA 16802, USA} \affiliation{Dept. of Physics, Pennsylvania State University, University Park, PA 16802, USA} \affiliation{Dept. of Physics and Astronomy, Uppsala University, Box 516, S-75120 Uppsala, Sweden} \affiliation{Dept. of Physics, University of Wuppertal, D-42119 Wuppertal, Germany} \affiliation{DESY, D-15738 Zeuthen, Germany}  \author{R. Abbasi} \affiliation{Department of Physics, Loyola University Chicago, Chicago, IL 60660, USA} \author{M. Ackermann} \affiliation{DESY, D-15738 Zeuthen, Germany} \author{J. Adams} \affiliation{Dept. of Physics and Astronomy, University of Canterbury, Private Bag 4800, Christchurch, New Zealand} \author{J. A. Aguilar} \affiliation{Universit{\'e} Libre de Bruxelles, Science Faculty CP230, B-1050 Brussels, Belgium} \author{M. Ahlers} \affiliation{Niels Bohr Institute, University of Copenhagen, DK-2100 Copenhagen, Denmark} \author{M. Ahrens} \affiliation{Oskar Klein Centre and Dept. of Physics, Stockholm University, SE-10691 Stockholm, Sweden} \author{C. Alispach} \affiliation{D{\'e}partement de physique nucl{\'e}aire et corpusculaire, Universit{\'e} de Gen{\`e}ve, CH-1211 Gen{\`e}ve, Switzerland} \author{A. A. Alves Jr.} \affiliation{Karlsruhe Institute of Technology, Institute for Astroparticle Physics, D-76021 Karlsruhe, Germany } \author{N. M. Amin} \affiliation{Bartol Research Institute and Dept. of Physics and Astronomy, University of Delaware, Newark, DE 19716, USA} \author{R. An} \affiliation{Department of Physics and Laboratory for Particle Physics and Cosmology, Harvard University, Cambridge, MA 02138, USA}
% ...
\date{\today}  
 \collaboration{IceCube Collaboration} 
 \noaffiliation
\date{\today}

\begin{abstract}
Solar flares convert magnetic energy into thermal and non-thermal plasma energy, the latter implying particle acceleration of charged particles such as protons. Protons are injected out of the coronal acceleration region and can interact with dense plasma in the lower solar atmosphere, producing mesons that subsequently decay into gamma rays and neutrinos at $\mathcal{O}$(MeV-GeV) energies. We present the results of the first search for GeV neutrinos emitted
during solar flares carried out with the IceCube Neutrino Observatory. While the experiment was originally designed to detect neutrinos with energies between 10~GeV and a few PeV, a new approach allowing for a $\mathcal{O}$(GeV) energy threshold will be presented. The resulting limits allow us to constrain some of the theoretical estimates of the expected neutrino flux. 
\end{abstract}
\maketitle
\section{Introduction}

While multi-messenger astronomy has recently recorded major breakthroughs such as the first joint observations of high-energy neutrinos and electromagnetic radiation from a single source~\cite{TXS}, the interior and exterior of the Sun have been detected through several messengers for decades.
Both the quiescent and the active phases of the corona are studied through the electromagnetic radiation emitted across the entire spectrum. 
However, in the neutrino search presented in this paper, we only focus on the gamma rays produced by high-energy protons that are accelerated in solar flares and subsequently collide with dense layers of the solar atmosphere. These gamma rays can be detected, e.g. by the Fermi-LAT satellite~\cite{fermi-lat-10-years, fermicatalog}, which significantly increased the fraction of solar flares detected in the high-energy range.

So far, solar neutrinos have only been detected in the MeV range being produced by fusion reactions in the core of the Sun. Large neutrino telescopes have recently started to search for solar atmospheric neutrinos~\cite{icecube-solaratm}, which are created by the collisions of high-energy cosmic rays colliding with the solar atmosphere~\cite{solaratm1, solaratm2, solaratm3, solaratm4}. While only upper limits have been established so far, the observation of this flux would allow us to probe solar magnetic field topology and would constitute a neutrino floor for dark matter searches in the core of the Sun. 
The same high-energy cosmic-ray collisions with the solar atmosphere create a muon shadow (deficit) in the Sun's direction. As a fraction of the cosmic rays are absorbed in the solar atmosphere or deflected by the solar magnetic field, the flux reaching the Earth atmosphere is reduced and so is the subsequent secondary atmospheric muon flux detected by IceCube~\cite{sunshadow}.  The variations in the shadow can be used to constrain the solar coronal magnetic field~\cite{sunshadow_theory}.

In addition to the aforementioned emissions that are expected to be continuous in time, transient neutrino emissions could be produced by solar flares.
Despite several searches since the late eighties~\cite{homestake, bahcall, kamiokande, sno}, no significant neutrino signal from solar flares has been confirmed to date. 
The main interest in searching for solar flare neutrinos comes from their hadronic origin: being inherent products of high-energy proton collisions in the chromosphere, neutrinos represent a direct probe of the proton acceleration. Theoretical investigation \cite{moriondproceeding, thesis, fargion} have demonstrated that this neutrino flux could extend from MeV up to a few GeV in energy. Focusing on the high-energy component of the solar flare neutrino spectrum would allow us to probe the proton acceleration up to the highest energies that can be reached within the solar flare environment.

This paper describes first the scientific motivation for and the associated production mechanism of the GeV neutrinos in solar flares. The IceCube Neutrino Observatory used for this search was built to detect neutrinos with energies larger than 10~GeV.  A new analysis approach developed for this study allows for the measurement of neutrinos at the single GeV energy scale. In Section~\ref{section:selection}, we present the scheme of how we select a special class of solar flares for this study based on the tight link existing between gamma rays and neutrinos. The IceCube Neutrino Observatory used for this neutrino search is described in Section~\ref{section:icecube} and the event selection we have developed is presented in detail in Section~\ref{section:eventselection}. Finally, the first results of a solar flare neutrino search carried out with IceCube are presented in Section~\ref{section:results}, along with perspectives for the next solar cycle. 

\begin{figure*}
	\centering
	\includegraphics[width=0.75\linewidth]{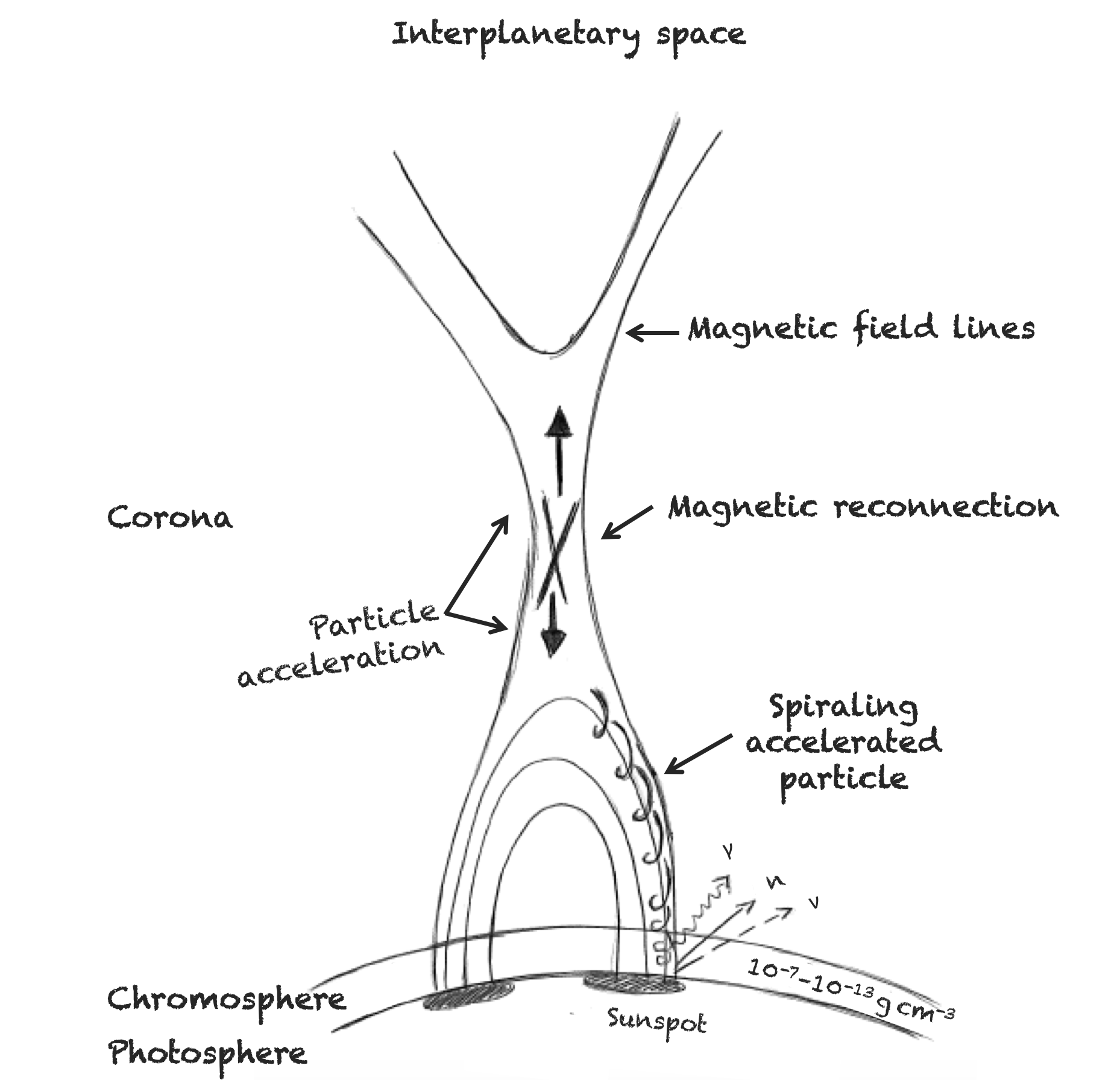}
	\caption{Schematic representation of a solar flare. The cross represents the magnetic reconnection and the two arrows show the direction of the subsequent outflow jets.\label{flare}}
 \end{figure*}

\section{The Potential Physical Constraints Brought by Neutrinos}
Solar flares convert magnetic energy into plasma heating and kinetic energy of charged particles such as protons~\cite{hudson}. As illustrated in Fig.~\ref{flare}, protons are ejected away from the coronal acceleration region and interact with the dense plasma in the lower solar atmosphere, producing neutrinos through the following reactions, 
\begin{widetext}
\begin{equation}
p\, +\,  p \,{~\rm or~} \, p\, + \, n
\longrightarrow 
\left\{
\begin{array}{l}
 \pi^+ + X; \\
  \pi^0 \,+ X; \\
    \pi^- + X; \\
    \end{array}
\right.\\
\label{reaction}
\\
\begin{array}{l}
\pi^+ \longrightarrow \mu^+ + \nu_{\mu} ;~ \mu^+ \longrightarrow e^+ + \nu_e + \bar\nu_{\mu} \\
\pi^0 \longrightarrow 2 \gamma \\
\pi^- \longrightarrow \mu^- + \bar{\nu}_{\mu}; ~\mu^- \longrightarrow e^- + \bar{\nu}_e + \nu_{\mu}. \\
\end{array}
\end{equation}
\end{widetext}
Here, the kinetic energy threshold for this process is 280 MeV for both proton-proton and proton-neutron interactions.

For the interpretation of our result, we assume that the accelerated proton flux can be modeled by the functional form $d\phi/dE = A E^{-\delta} H(E_{\max} - E)$, where $A$ is a normalization constant, $\delta$ represents the spectral index, and $E$$_{\max}$ is the upper cutoff in a Heaviside function, as motivated in~\cite{trottet}.  

\begin{figure}
	\centering
	\includegraphics[width=0.65\textwidth]{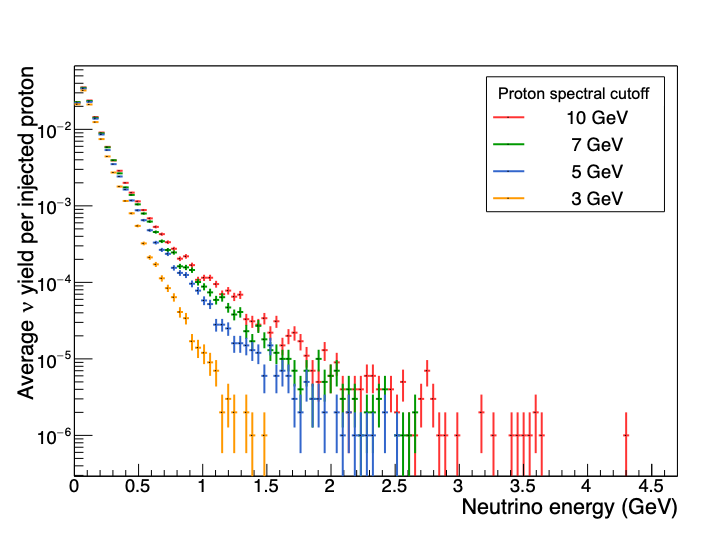}
	\caption{Energy distribution for neutrinos depending on the upper cutoff in the
accelerated proton spectrum with a spectral index $\delta=3$. The distributions have been obtained using a GEANT4-based simulation of proton-proton interactions in the chromosphere.\label{upper_cutoff}}
 \end{figure}

The proton spectral index has been extracted fitting gamma-ray observations by Fermi-LAT and assuming a pion decay model for different phases of the solar flare. The data were provided by the Fermi-LAT Collaboration based on an analysis similar to the one carried out in ~\cite{fermilde}. So far, there are no constraints on the value of the upper cutoff. The effect of this upper cutoff on the subsequent neutrino flux is illustrated in Fig.~\ref{upper_cutoff}, where the colored points show the average neutrino yield per injected proton when assuming an initial proton flux following a power law with a spectral index $\delta$ = 3 and realistic values of $E_\mathrm{max}$ in the $(3-10)$ GeV region. This result is obtained using a GEANT4-based simulation of high-energy proton collisions with the solar atmosphere. The accelerated protons were injected in a direction tangent to the photosphere as it leads to a better agreement with gamma-ray observations by Fermi-LAT.  Most of the neutrinos are produced in the Chromosphere at densities around 10$^{-7}$ - 10$^{-8}$ g cm$^{-3}$ in our simulation. More details about the simulation can be found in \cite{thesis}.
As it can be seen in Fig~\ref{upper_cutoff}, a higher cutoff value leads to a higher neutrino yield in the GeV energy range and would thus lead to a larger signal in sensitive neutrino telescopes.
Coupling Fermi-LAT and IceCube observations has therefore the potential to constrain both this upper cutoff and the spectral index by measuring the strength of the detected neutrino signal in IceCube and the fitted 
spectral index of the gamma ray spectrum.

\section{Selection of Solar Flares}\label{section:selection}

Following Eq.~(\ref{reaction}), one finds that pion production generates both neutrino and gamma-ray emissions. We focus only on solar flares that must emit neutrinos at some level and use the gamma-ray observations to pick the most relevant candidates for our neutrino search. 
These are indications that pion-decay products dominate the gamma-ray spectrum above 100 MeV~\cite{vilmer}, an energy that Fermi-LAT has the ability to detect. A spectral analysis confirms that the major contribution of the Fermi-LAT observations was indeed consistent with pion decay emissions~\cite{ackermann}. 

We assume that the neutrino emission is coincident in time with significant pion decay signals detected in Fermi-LAT during solar flares. This leads to a significant difference to previous solar flare samples~\cite{kamiokande, sno} that were based on the X-ray flux. For example, Fermi-LAT only detects 5\% of the M- and X-solar flares, i.e. solar flares with a X-ray brightness in the wavelength range between 1 and 8 {\AA}ngstr\"oms above 10$^{-5}$ and 10$^{-4}$ W/m$^2$, respectively, detected by the Geo-Stationary Operational Environmental Satellite (GOES) on average~\footnote{This number has been obtained by counting the number of solar flares detected by Fermi-LAT between 2011 and 2015 and the number of M and X-class flares. The field of view of Fermi-LAT allows for solar observations during 20\% to 60\% of the time~\cite{fermi-LAT-general} and cannot explain by itself this small fraction of observations.}~\cite{goes}.

\begin{figure}
	\centering
	\includegraphics[width=0.75\textwidth]{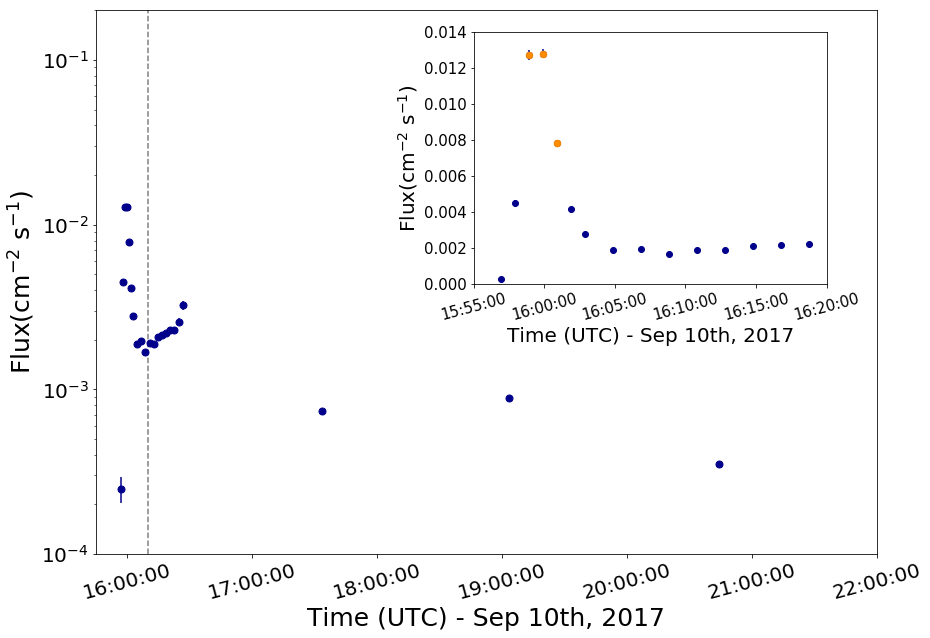}
	\caption{Gamma-ray light curve of a solar flare recorded above 100~MeV on Sept 10th, 2017 by Fermi-LAT (main plot) and selected time window for the neutrino search (orange points in the inset). The dashed line distinguishes the impulsive phase from the long duration emission. The data points are kindly provided by the Fermi-LAT Collaboration.  \label{sep10_selection}}
 \end{figure}

The light curve recorded by X-ray devices and by Fermi-LAT usually shows a short high-intensity peak on top of a lower baseline flux with an underlying period~\cite{vilmer}.  An example of such a flare is shown in Fig.~\ref{sep10_selection}.  The peak is referred to as the \textit{impulsive phase}. The analysis of the gamma rays detected during these impulsive short phases reveals a relatively hard initial proton spectrum, with a spectral index around 3~\cite{fermilde}.
In contrast, the long duration emissions manifest themselves in a softer proton spectral index (typically between 4 and 6) and a spread of the gamma-ray emission over several hours.
Focusing on the impulsive phase of bright events of the 24$^{\text{th}}$ solar cycle allows one to minimize the  background integrated in the neutrino telescope and to thus increase the chance of a neutrino detection in coincidence with solar flares.

These criteria applied to the first Fermi-LAT Solar flare catalog~\cite{fermicatalog} resulted in a list of 5 promising candidates for our neutrino search. The details of each analyzed solar flare are reported in Table~\ref{table:timewindow}. The choice of the time window and duration of each solar flare was made in view of maximizing the signal-to-noise ratio in IceCube. We started from the maximum flux recorded by Fermi-LAT and kept integrating until our signal-to-noise ratio started to decrease. As an example, the result of the time window selection for the solar flare of Sept 10th, 2017 is shown by the orange points in Fig.~\ref{sep10_selection}. This optimal time window represents 45\% of the observation time window reported by Fermi-LAT. The observed fraction of the other solar flare events considered in this work are reported in Table~\ref{table:timewindow}.

\begin{table}
\centering
\caption{Optimized time window for neutrino searches. We indicate the position of the flare on the solar disk for completness.}
\begin{tabular}{c|c| c |c|c}
Date & Selected  & Duration & Fraction &Location on the  \\
 & time window & (minutes) &observed &solar disk\\

\hline
March 7th, 2012  & 00:41:22 - 01:21:22 & 40 & 85\% &Centered, North-East quarter\\
February 25th, 2014  & 01:07:30 - 01:32:30& 25 & 97\% & Limb, South-East quarter\\
September 1st, 2014 & 11:07:00 - 11:21:00&  14 &95.5\% & 36${\degree}$ behind the East limb\\
September 6th, 2017 & 13:23:03  - 22:00:37 & 515&87\% & Centered, South-West quarter\\
September 10th, 2017 & 15:58:54 - 16:02:52& 5.96 &45\% & Limb, South-West quarter\\
\end{tabular}
\label{table:timewindow}
\end{table}

\begin{figure*}
	\centering
	\includegraphics[width=0.7\linewidth]{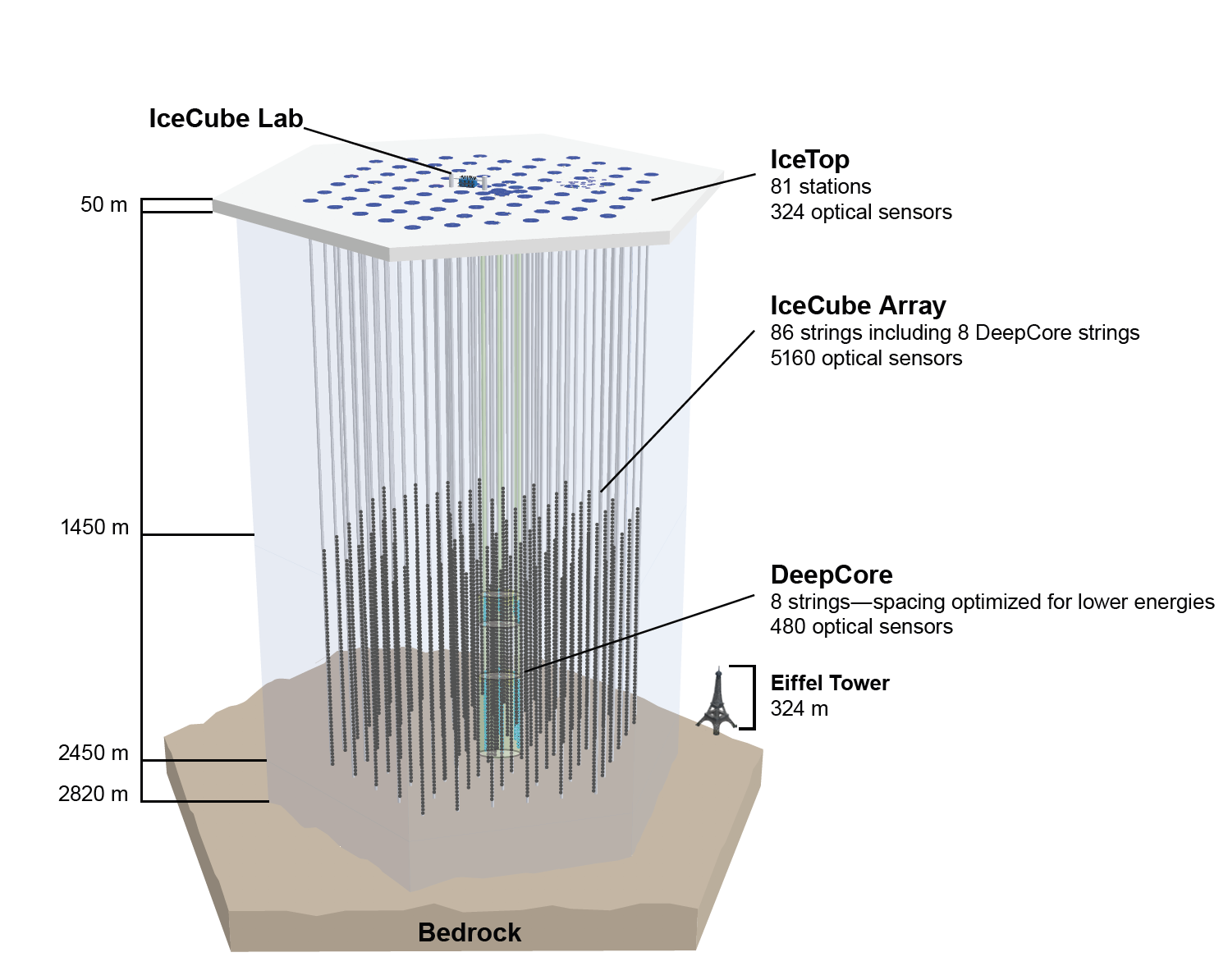}
	\caption{Schematic view of the IceCube Neutrino Observatory with the IceTop surface array and the DeepCore subdetector.  \label{icecube}}
 \end{figure*}

\section{The IceCube Neutrino Observatory and the search for GeV neutrinos}\label{section:icecube}\
The IceCube Neutrino Observatory consists of a cubic kilometer of instrumented ice located at the South Pole, 2~km below the surface~\cite{jinst}. Completed in December 2010, the detector is made of 86 vertical strings, with 60 digital optical modules, or DOMs, each. The strings have an average horizontal spacing of 125~m while the DOMs are located every 17~m along the instrumented portion of the string. A sub-detector, named DeepCore, is installed at the center of the array as shown in Fig.~\ref{icecube}. Characterized by smaller spacings between strings ($\approx$ 70~m) and DOMs ($\approx$ 7~m), DeepCore offers a lower energy threshold for neutrino detection, down to 10~GeV~\cite{deepcore}. No DOMs were deployed between depths of 2000~m to 2100~m where the optical scattering and absorption are significantly increased due to a dust layer~\cite{dustlayer}.
More details about the DOMs used as detection units can be found in~\cite{jinst} and references therein.

In order to reduce the noise rate, trigger conditions based on coincidences are applied. Different coincidence criteria classify the signal detected in IceCube~\cite{timestamping}: Hard Local Coincidence (HLC), when two neighbor or next-to-neighbor DOMs on the same string record a signal above threshold within a 1~$\mu$s time window; and Soft Local Coincidence when the hit does not qualify for the HLC criteria.  The main trigger for IceCube events is the Simple Majority Trigger, or SMT-8, which requires at least 8 DOMs with HLC pulses within a 5~$\mu$s time window. A softer trigger condition, SMT-3, has been implemented for DeepCore events in order to lower the energy threshold. It requires, in analogy to its counterpart in the full IceCube array, 3 DOMs with HLC pulses within 2.5~$\mu$s. Considering the low energy expected for solar flare neutrinos, we focus on events that have fulfilled the SMT-3 condition.

A parallel data stream allows IceCube to be sensitive to Galactic core-collapse supernovae~\cite{supernova}. Individual neutrino interactions in the MeV range produce a signal too dim to identify individual events, but the large flux emitted during a close-by supernova would lead to a detectable coherent rise in the individual hit rates of the DOMs. In this paper, we apply the principle used for MeV core collapse supernova neutrinos to the GeV energy range in order to search for neutrinos from solar flares. This is explained in more detail in the next section.
\\


\begin{figure*}[h!t]
\centering
\subfloat[PeV neutrino interaction]{\includegraphics[width=0.3\textwidth]{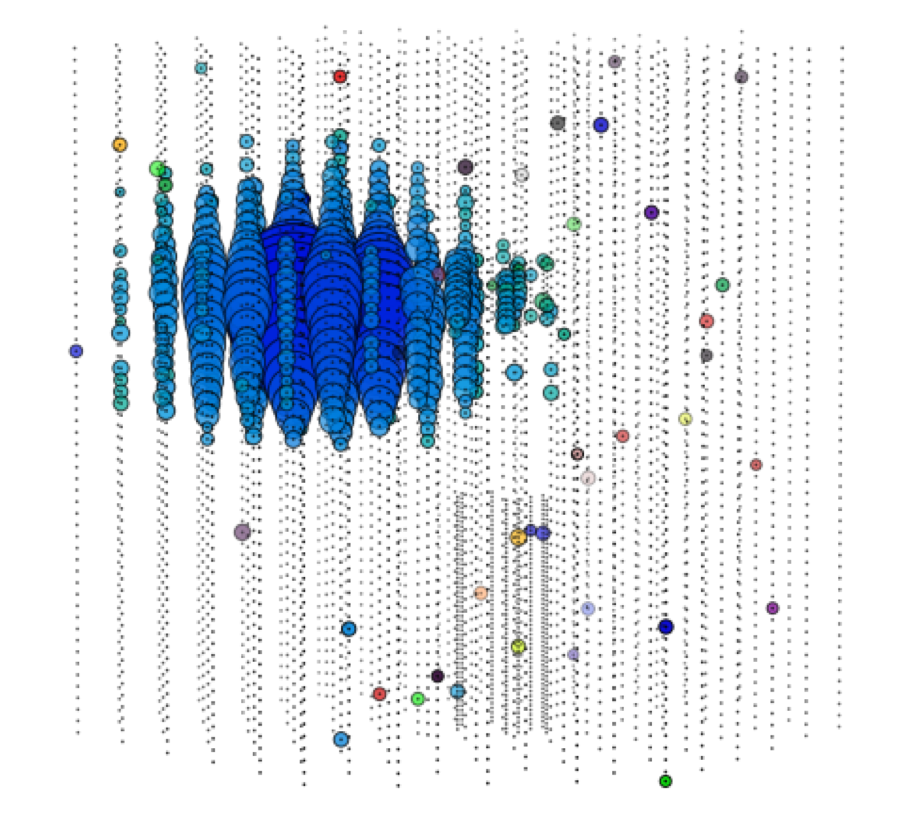}
            \label{hese}}
\subfloat[10 GeV neutrino interaction]{\includegraphics[width=0.3\textwidth]{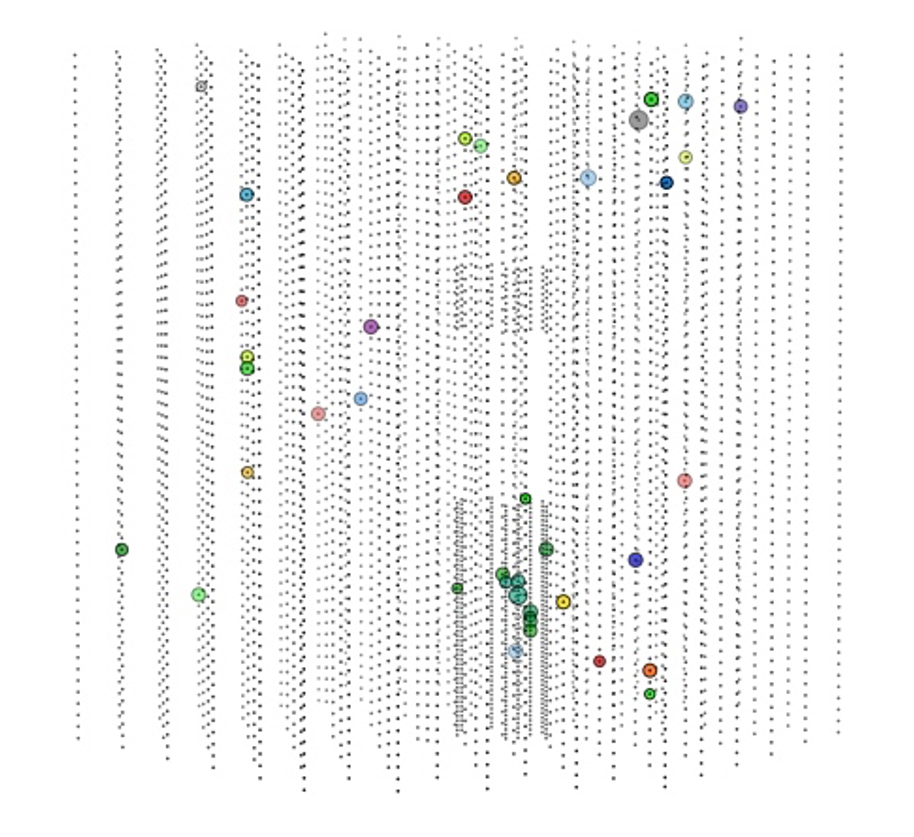}
            \label{hese}}
\subfloat[Detector noise event]{\includegraphics[width=0.3\textwidth]{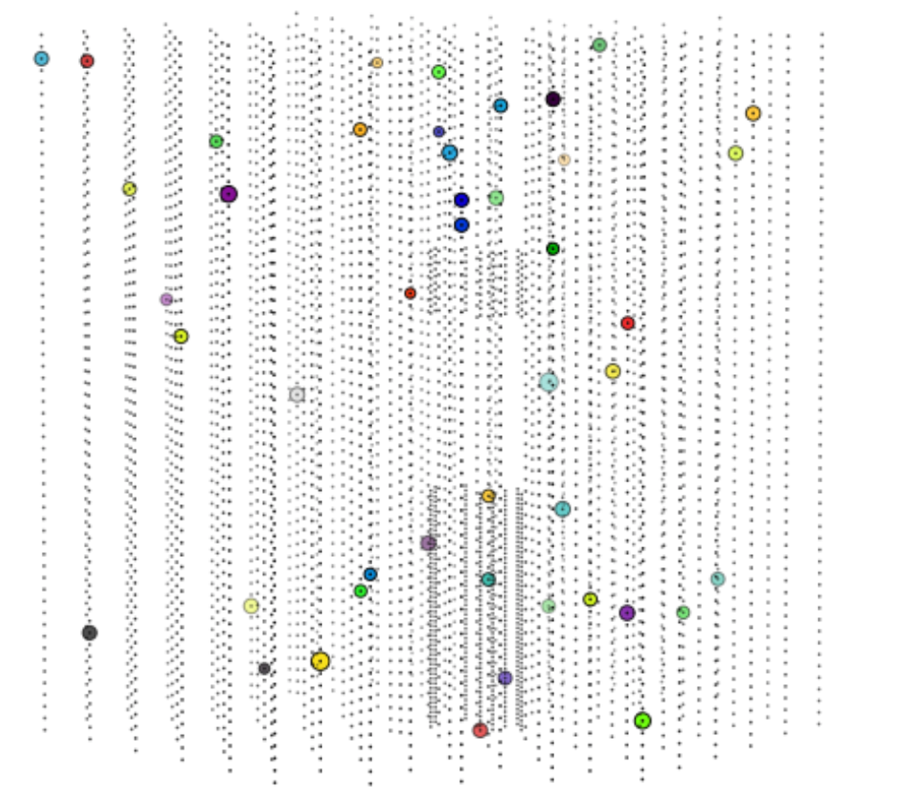}
            \label{hese}}            
            \hspace{.25in}%
\subfloat[GeV neutrino interaction]{\includegraphics[width=0.3\textwidth]{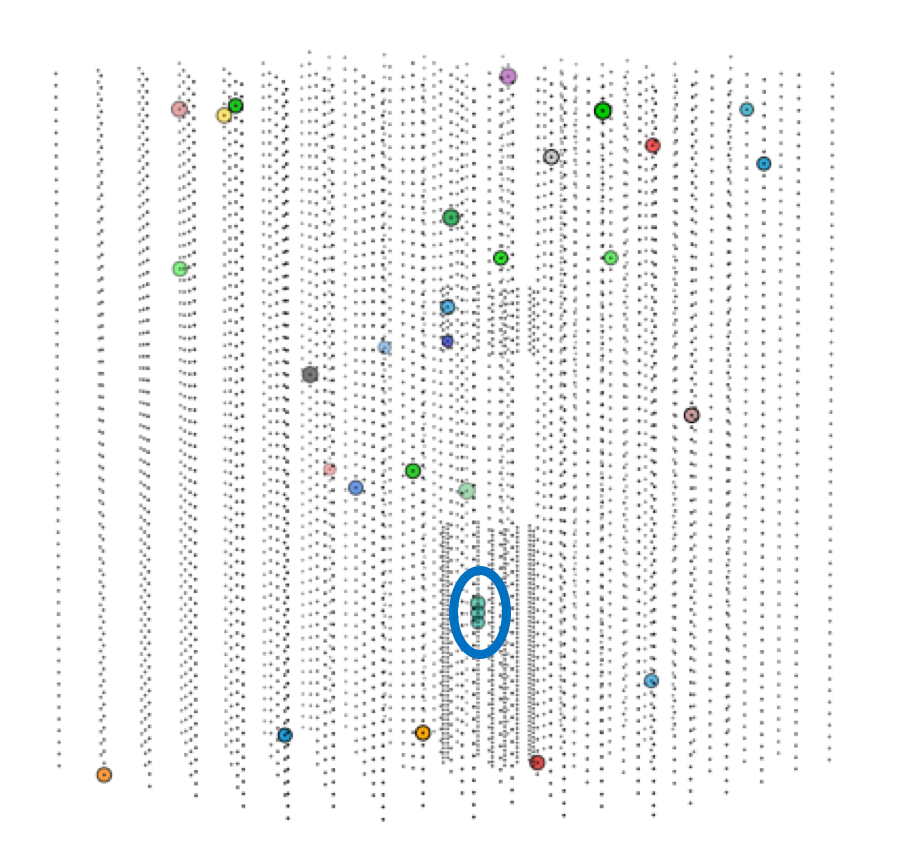}
            \label{sf}}
\subfloat[Focus on the GeV neutrino interaction]{\includegraphics[width=0.3\textwidth]{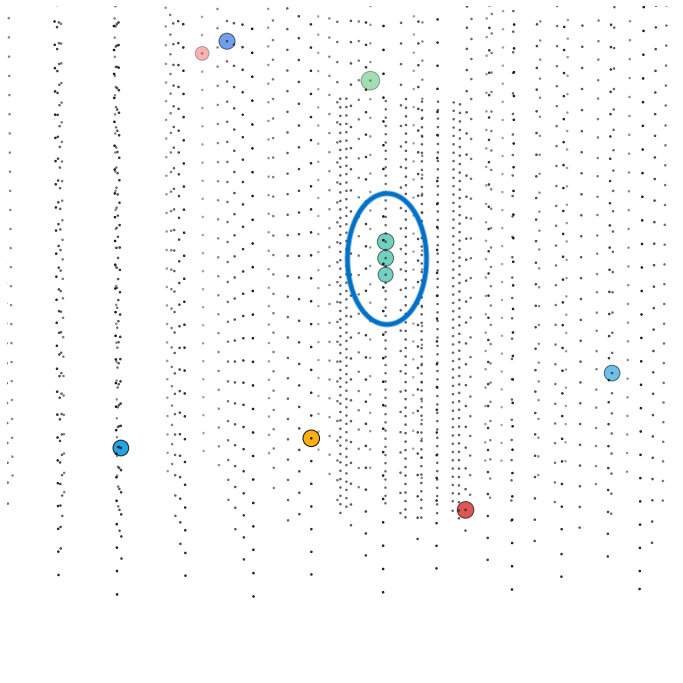}
            \label{sf}}            
            \hspace{.25in}%
 \caption{Examples of neutrino interactions as seen in IceCube. A typical GeV neutrino interaction is illustrated in (d)~(simulation, with a focus on the neutrino interaction in (e)) while (a)~(data) (b), and (c)~(simulation), show a well-known high-energy starting event with PeV energy, a typical 10 GeV neutrino interaction, and an event triggered by detector noise, respectively (see the text for more details). The colored points represent the DOMs that have recorded a signal during a time window of about 12~$\mu$s around the event. The size of the points represent the intensity of Cherenkov light that was detected. 
              }\label{fig:events}

\end{figure*}
\section{Optimized Search for GeV Neutrino Interactions}\label{section:eventselection}

Before this analysis was developed, neutrino searches in IceCube were possible in the MeV range by the use of the Supernova Data Acquisition system and starting at around 10~GeV for neutrino oscillation studies in DeepCore~\cite{jinst}. We introduce a novel selection procedure that enables us to cover part of the gap between these two energy ranges. 
The general idea of the analysis is to monitor the rate of GeV-like events in IceCube and to search for an increase in this rate during an astrophysical transient, as for example a solar flare. While the principle is similar to the one in use for MeV supernova neutrino searches, the neutrino flux produced by solar flares is expected to be several orders of magnitude below the one of Galactic core-collapse supernovae. The Supernova Data Acquisition system could not be used for our purpose; we focus on events that have triggered the data taking in IceCube.

The new event selection has been developed in a data-driven way following a blind procedure. We have used several hours of IceCube data recorded when no solar flare had been detected on either side of the solar disk. In order to determine the time period when such data were available, we compared the observations by the X-ray instrument GOES~\cite{goes} and Fermi-LAT~\cite{fermi-lat-10-years}, which is sensitive to gamma rays. We have also used the observations of the STEREO-A and STEREO-B satellites that observe the hidden side of the Sun and detect, among others, energetic particles arising from the Sun~\cite{stereo}. In addition to these off-time data, we have simulated various classes of events to understand the behavior of the selection to different types of events present in data. We use CORSIKA~\cite{corsika} and GENIE~\cite{genie} to simulate atmospheric muons and neutrinos respectively, and a noise event generator based on an empirical model consisting of three noise components inherent to IceCube DOMs: uncorrelated thermal noise, uncorrelated radioactive noise, and correlated scintillation noise. Finally, signal-like neutrino interactions were generated with GENIE by selecting events 
with an energy between 500~MeV and 5~GeV that arrive from the observed declination range of the Sun at the South Pole, i.e. [-23, 23] degrees. We note that various neutrino interaction processes are included in the simulation - deep inelastic scattering, quasi-elastic scattering, resonant production - which all contribute substantially to the total interaction rate, as well as scattering that contributes only marginally.

The event selection is discussed in detail below. It can be divided into three main steps: removal of high-energy events from the sample, reduction of the contribution from events triggered by noise, and the increase of the purity of the final sample. 

\subsection{Removing High-Energy Events}
The first step of the event selection aims at reducing the number of high-energy events, in this work understood as neutrinos with an energy $\gtrsim$ 5~GeV, from the data sample because we expect solar flares to produce neutrinos only up to a few GeV. The main difference between an event with arbitrarily high energy and a GeV neutrino interaction is the amount of light emitted in the ice, as shown in Fig.~\ref{fig:events}. We use available data streams (``software filters") that have been developed to tag specific kind of events such as high energy muons or cascades. As depicted in Fig.~\ref{fig:events}(d), our events activate a small number of DOMs and are thus not expected to pass any of the filters designed to tag high-energy interactions. To be part of our event sample, an event therefore has to pass the filter that selects events contained in DeepCore and fail all other filter conditions. An exception is made for two filters: one targeting low-energy neutrinos coming from the Northern sky, and the one that uses parts of the detector as veto against incoming muon events, as both subsequent samples contain low-energy events~\cite{jinst}. This combination of filters results in a significant reduction of the number of atmospheric muons: the event rate after applying this filter selection is of the order of 15~Hz while the original rate was around 1400~Hz. More than 98\% of the simulated neutrino events between 500~MeV and 5 GeV pass this filter selection.\\
The number of HLC hits in IceCube and DeepCore strings is small for low-energy neutrino events, while background events from atmospheric muons and neutrinos are typically characterized by a large number of HLC hits in the detector. We can therefore impose a constraint on the maximum number of HLC hits allowed in DeepCore and the strings outside DeepCore to remove high-energy events from our sample. More stringent constraints on the number of HLC hits in IceCube strings do not improve the signal over noise ratio of the selection because of noise hits on the considered strings present in both signal and background events.
The number of hits that share a causal connection inside an event can also be used as a parameter for the amount of light emitted in the detector as a consequence of the neutrino interaction. We use an algorithm designed to select sets of hits most likely connected to the same physical interaction and therefore unrelated to dark noise. The algorithm selects hits that have at least one other hit within a sphere of radius $R$ = 150~m and within a time window of $\Delta T$ = 1000~ns called Single R-T hits, or SRT hits hereafter. The constraints listed in Table~\ref{tab:sum-tot} lower the data rate to about 5.5~Hz while keeping 98\% of the simulated signal events in the sample.

\subsection{Minimizing the Contribution of Pure Noise}
Noise triggers occur with relatively few hits and no preferred direction. Figure~\ref{fig:events}(c) represents a typical noise event satisfying the SMT-3 trigger condition and passing the filter selection previously described. Accidental triggers by pure noise constitute a significant background for low-energy interactions such as the ones illustrated in Fig.~\ref{fig:events}(d). The IceCube Collaboration has developed an algorithm able to identify and eliminate noise signals. The algorithm searches for a preferential direction in pairs of hits to classify the event as being of physics origin. In practice, an event is classified as \textit{physics} if it contains, during a certain time window ($W$), a minimum of $N$ pairs of hits with an effective particle velocity contained in a [$v_{min}$,$v_{max}$] m/ns interval pointing in excess towards a certain direction. If this is not the case, the event is classified as \textit{noise}. 

The default set of these four parameters ([$W$, $N$, $v_{\min}$,$v_{\max}$] = [500~s, 7, 0.05~m/ns, 0.5~m/ns]) is implemented in the IceCube software and regularly used for data analyses (see e.g.,~\cite{tau-appearance}). The algorithm has a Boolean output with TRUE attributed to physics-like events and a likely noise event corresponds to FALSE. This optimization allows one to differentiate noise from 10-100~GeV neutrino interactions. In order to be effective for our lower energy events, we have re-optimized the method.  We have applied the algorithm for each possible set of parameters and constructed combinations of these sets that maximize the signal to noise ratio. We request an output FALSE for sets of variables specifically targeting small apparent speed in a long time window (e.g., [800~s, 0 , 0.00~m/ns, 0.10~m/ns]) as these parameters correspond to scattered hits. 
On the contrary, the sets for which a TRUE output is required to pass the selection lead to a sample dominated by high apparent speed in a short time window, i.e. unscattered hits (e.g., [100~s, 2, 0.20~m/ns, 0.90~m/ns]). \\
The data rate is about 0.2~Hz after applying the two combinations of parameter sets shown in Table~\ref{tab:sum-tot}. For comparison, the noise rate is estimated to be around 0.12~Hz. More than 55\% of simulated signal events survive these cuts.

\subsection{Increasing the Purity}
With most of the noise events removed from the sample, we continue to increase its purity.
Several selection criteria are applied: 
\begin{itemize}
\item \textit{Charge distribution}: cut on the ratio of the charge deposited in DeepCore during the first 600~ns after the first HLC hit in DeepCore and the total charge in the event. This cut allows one to remove remaining low-energy tracks created by atmospheric muons that travel across the detector. To be kept in the sample, an event is requested to have a charge ratio $\leq$ 0.26.
\item \textit{Depth}: cut on the depth of the first HLC hit in DeepCore. Low-energy atmospheric muons will leave more energy in the top of the detector rather than in deeper DOMs. The optimal depth for the selection of GeV neutrino events has been defined as [-2453~m, -2158~m], which corresponds to the part of DeepCore located below the dust layer.
\item \textit{Centroid of the event}: cut on the distance and the time delay between the first and the second HLC hits in DeepCore. Since our events are of low energy, the DOMs able to record the events are
close to each other and the hits happen within a short time window. We requested the distance between these two hits to be smaller than 70~m with a time delay not more than 50~ns.
\item \textit{Total charge}: the total deposited charge inside the detector must be lower than 60 photoelectrons.
\end{itemize}

\begin{table*}
\begin{center}
\caption{Summary of the cuts applied in the event selection.}
\begin{tabular}{c| c l }
\hline
{\textbf{Variable}} & {\textbf{Passing conditions}} \\
\hline
\multicolumn{2}{c}{\textbf{Initial data rate: 1400~Hz}}\\
\hline
 &Events contained in DeepCore \\
 Passing filters &  and failing every other software filters\\
 & except two targeting low-energy events\\
 \hline
 \multicolumn{2}{c}{\textbf{Data rate: 15~Hz - Passing signal events: 98~\%}}\\
\hline
 Number of HLC in IceCube w/o DeepCore, $N_1$,  &  $N_1$ = 0 and $N_2$ $\leq$ 9\\
 and number of HLC in DeepCore, $N_2$  & OR $N_1$ $\leq$ 6 and $N_2$ $\leq$ 7 \\ 
 \hline
 Number of SRT hits &$\leq$ 10  \\
 \hline
 \multicolumn{2}{c}{\textbf{Data rate: 5.5~Hz - Passing signal events: 93~\%}}\\
\hline
Noise algorithm variables & [100~s, 2, 0.20~m/ns, 0.90~m/ns] = TRUE\\
Step 1 &   [100~s, 0, 0.20~m/ns, 0.90~m/ns] = TRUE \\ 
$\big[$$W$, $N,$ $v_{\min}$,$v_{\max}$$\big]$ & AND [1000~s, 0, 0.00~m/ns, 0.10~m/ns] = FALSE\\
\hline
Noise algorithm variables&[300~s, 2, 0.20~m/ns, 0.40~m/ns] = TRUE\\
Step 2 &  [300~s, 2, 0.10~m/ns, 0.90~m/ns] = TRUE \\
$\big[$$W$, $N$, $v_{\min}$,$v_{\max}$$\big]$  & AND [800~s, 0, 0.00~m/ns, 0.10~m/ns] = FALSE \\
&   [500~s, 2, 0.20~m/ns, 0.30~m/ns] = TRUE\\
\hline
 \multicolumn{2}{c}{\textbf{Data rate: 0.2~Hz - Passing signal events: $>$ 55~\%}}\\
\hline
Charge ratio & $>$ 0.26 \\
\hline
Depth of the first HLC in DeepCore & [-2453 m, -2158 m]\\
 \hline
Distance and delay between & $<$ 70~m  \\
 1st and 2nd HLC in DeepCore&  $<$ 50~ns\\
 \hline
 \hline
Total charge & $<$ 60 photoelectrons \\
 \hline
 \multicolumn{2}{c}{\textbf{Data rate: 0.02~Hz - Passing signal events:$ >$ 35~\%}}\\
 \hline
\end{tabular}
\label{tab:sum-tot}
\end{center}
 \end{table*}

A summary of the cuts applied for the final selection is presented in Table~\ref{tab:sum-tot}. The data rate, constant over time, is 20 $\pm$ 2~mHz, which can be compared with the simulated rates of 18~mHz for pure noise events, less than 5~mHz  for atmospheric muons, 0.3~mHz and 0.8~mHz for $\nu_e$+ $\bar{\nu}_e$ and $\nu_\mu$+$\bar{\nu}_\mu$, respectively. A study of these off-time data showed that the number of events recorded in short time windows were consistent with a Poisson distribution.

The leading systematic uncertainties in this analysis arise from our limited understanding of the optical properties of the surrounding ice. We studied in particular the impact of the DOM efficiency and the scattering length in the ice directly surrounding the DOM. This is the ice in the approximately 60 cm diameter bore hole that refroze after the DOMs were deployed. The DOM efficiency impact study was carried out using a conservative variation of $\pm$ 10\% of the quantum efficiency of the photocathodes in the simulation of low-energy neutrino interactions. Two different scattering lengths (namely, 100~cm and 50~cm) were used in simulations to assess the corresponding effect on low-energy interaction detection.
We concluded that the uncertainty on each parameter would lead to a number of low energy events within 20\% of the nominal value. As a GeV neutrino interaction needs to happen close from a DOM to trigger the data taking, the uncertainties on the light absorption and scattering in the bulk ice will have a negligible impact compared to the two other systematic effects previously mentioned.

\begin{figure}[h!]
	\centering
	\includegraphics[width=0.65\textwidth]{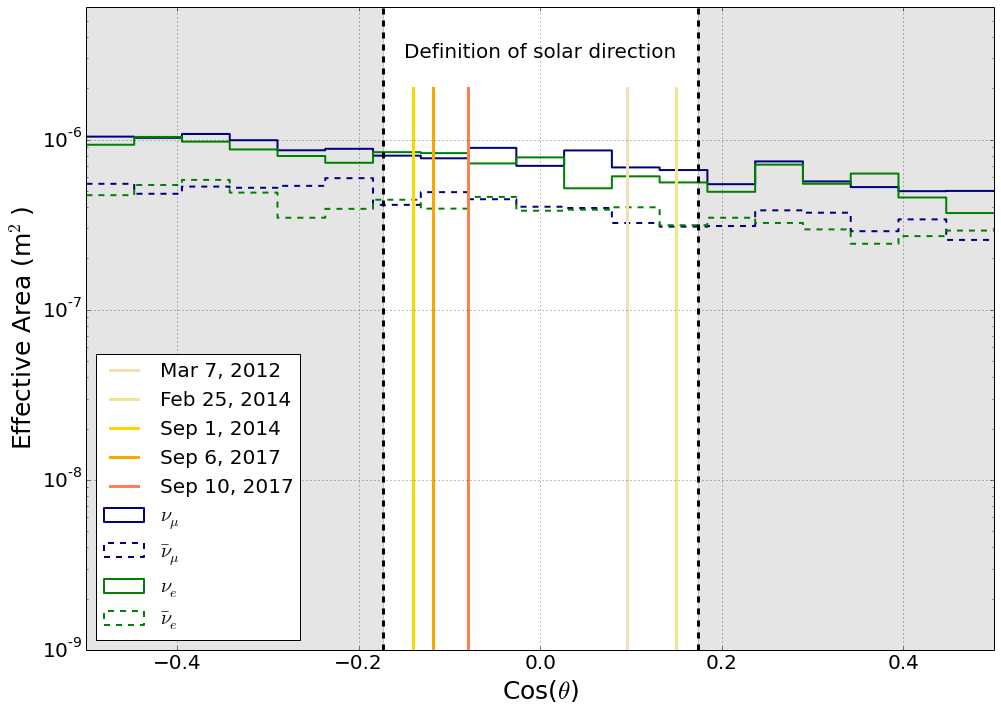}
	\caption{Effective areas for GeV-like neutrino events (500~MeV - 5~GeV) as function of their zenith coordinate. The colored vertical lines show the location of the solar flares studied in this work. The white region highlights the zenith range covered by the Sun at the South Pole.\label{fig:effectivearea}}
 \end{figure}

The passing rate of GeV neutrino events is of the order of 40~\% and shows only a small zenith dependence (see Fig.~\ref{fig:effectivearea}). Therefore, the event selection previously described can be used to search for transient events at an arbitrary position in the sky.
Fig.~\ref{fig:effectivearea} shows the effective areas for events as function of the zenith. We highlighted the position of the solar flares that are considered in this work by colored vertical lines.

\subsection{Searching for a statistical fluctuation}

We use the statistical test by Li and Ma~\cite{liandma} to quantify a possible excess of events. This method has been developed to estimate the significance of events in a certain time region, with the null-hypothesis defined as the non-existence of a signal source. Three parameters are used and their description in the framework of our analysis is:

\begin{itemize}
\item $N_\mathrm{on}$ the number of events in our final sample during the optimized solar flare time window.
\item $N_\mathrm{off}$: the number of events in our final sample during 8~hours prior to the solar flare onset. 
\item $\alpha$ = $t_{\text{on}}/t_{\text{off}}$, where $t_{\text{off}}$ is the duration of the time window considered prior to the solar flare, i.e. 8~hours, and $t_{on}$, the selected time window during the solar flare.
\end{itemize}

The estimate of the significance S proposed in~\cite{liandma} can be used under the conditions that the counts ($N_{\text{on}}$, $N_{\text{off}}$) were obtained by a single observation, where $N_{\text{on}}$ and $N_{\text{off}}$ are not too few. If these conditions are fulfilled, S refers to the number of standard deviations of the event  ($N_{\text{on}}$, $N_{\text{off}}$). This approach is particularly interesting when, as it is the case in this analysis, there are two unknown parameters: 
\begin{itemize}
\item $N_{\text{s}}$: the number of signal events, equivalent to $N_{\text{on}}$ - $\alpha$ $N_{\text{off}}$
\item $N_{\text{B}}$: the number of background events, during the time window of the search that is now precisely known.
\end{itemize}


\section{Results and Outlook}\label{section:results}

Table~\ref{tab:resultsnumberofevents} shows the number of off-source and on-source IceCube events as well as the corresponding significance $S$, calculated using Eq.~\ref{eq:liandma} following the Li and Ma approach~\cite{liandma}

\begin{widetext}
\begin{equation}
S = \sqrt{2} \left( N_{\text{on}} \; \ln \left[ \frac{1 + \alpha}{\alpha} \left( \frac{N_{\text{on}}}{N_{\text{on}} + N_{\text{off}}}\right)\right]  + N_{\text{off}} \; \ln \left[ (1 + \alpha) \left( \frac{N_{\text{off}}}{N_{\text{on}} + N_{\text{off}}}\right) \right]\right)^{1/2}.
\label{eq:liandma}
\end{equation}
\end{widetext}
We observe no significant signal. We can therefore derive upper limits on the potential number of signal events  solving $N_{\text{s}}$ = $N_{\text{on}}$ - $\alpha$  $N_{\text{off}}$ and using Eq.~\ref{eq:aeff}, which links the number of observed events and the neutrino fluence $\Phi$: 
\begin{equation} N_{\text{s}} = \int A_{\text{eff}}(E) \; \Phi(E) \; dE,\label{eq:aeff}\end{equation}
where A$_{\text{eff}}$ is the effective area shown in Fig.~\ref{fig:effectivearea}.
The final upper limits presented in Table~\ref{tab:resultsnumberofevents} take into account the flavor ratio at Earth after oscillations, assumed as $\nu_e$ : $\nu_{\mu}$ : $\nu_{\tau}$ = 1 : 1 : 1, and the effective area expected for each flavour and interaction type. \\
As shown in Table~\ref{table:timewindow}, the optimized time window for our neutrino search only contains a fraction of the observed gamma-ray light curve. Since this analysis assumes that the neutrinos are emitted jointly with the gamma rays, we assume our time windows contain the same fractions of the total neutrino emission. We therefore take these fractions into account when comparing with theoretical predictions. The limits presented in Table~\ref{tab:resultsnumberofevents} constrain the integrated neutrino flux emitted during the considered time window.

We note that similar number of events and fractions of the light curves are observed for the events of March 7th, February 25th and September 1st (see Table~\ref{tab:resultsnumberofevents}). This is reflected in the similar upper limits set for these events. Approximatively half of this fraction is contained in the short flare of September 10th, which explains the apparent lower upper limit. Finally, the September 6th flare targets the long duration emission that lasted for several hours, as previously mentioned. The origin of a such temporally extended emission is still under investigation~\cite{fermicatalog}, and therefore cannot, at the moment, be compared with the neutrino upper limits set for the other solar flares.

\begin{figure*}[h!t]
	\centering
	\includegraphics[width=0.65\textwidth]{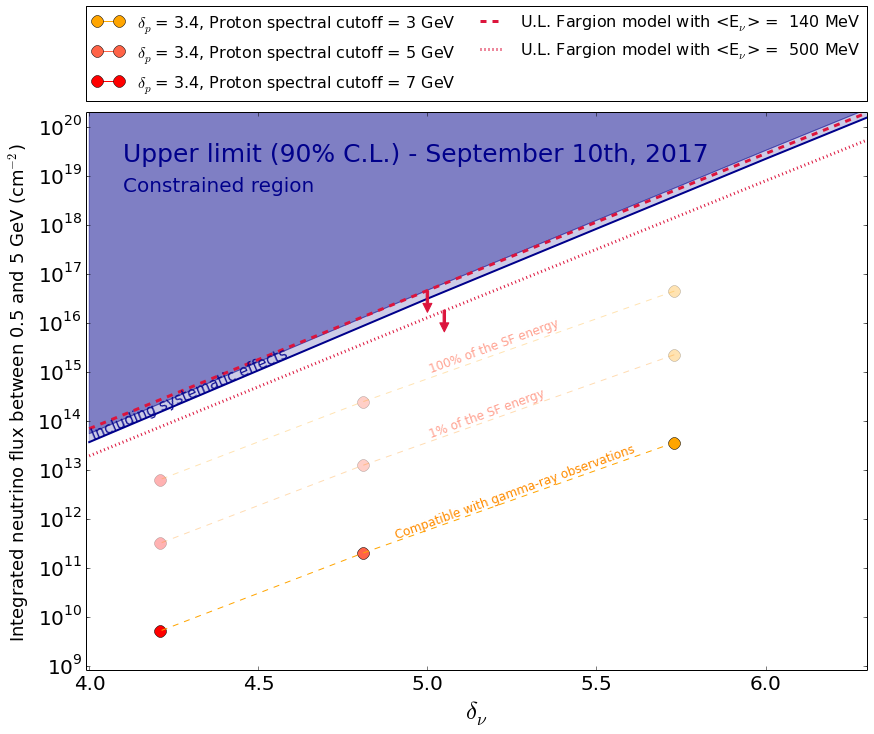} 
	\caption{Comparison of the experimental upper limit derived for the September 10,
2017 solar flare and the corresponding theoretical predictions. The orange points are derived from~\cite{thesis}, using a simulation assuming a proton spectral index of 3.4, derived from gamma-ray observations. The three sets of points are obtained assuming a different amount of energy going to accelerated protons, namely 10$^{32}$erg (100\% of the maximum magnetic energy available is converted into accelerated protons), 10$^{30}$erg (1\% of the magnetic energy is converted into accelerated proton), and 10$^{29}$erg, which is compatible with gamma-ray observations. The red line shows the predictions from~\cite{fargion}, with $E_\mathrm{flare}=10^{32}$ erg and
$\left < E_{\nu_e} \right>=140$ MeV (dashed) and 500 MeV (dotted). \label{UL}}
 \end{figure*}

The exact spectral index $\delta_{\nu}$ of the neutrino spectrum is not precisely known but could be estimated to lie between 4 and 6 using a GEANT4~\cite{geant4} simulation of proton-nucleus interaction in a solar environment 
as described in~\cite{thesis}. The neutrino upper limits are presented as a function of the parameter space $(\delta_\nu,\,C)$, where $C$ is the integrated neutrino flux between 500~MeV and 5~GeV. The upper limit obtained from the Sept 10th, 2017 event is shown in Fig.~\ref{UL} together with its systematic uncertainties. 
In Fig.~\ref{UL}, the obtained upper limit is compared to two predictions~\cite{thesis,fargion}. The difference between these two predictions comes from different assumptions on the proton flux accelerated by the solar flare: the total energy released by the magnetic reconnection is converted into protons that subsequently produce pions producing neutrinos is considered in~\cite{fargion}, while only a small fraction of the released energy, consistent with the estimates based on Fermi-LAT observations, is used in~\cite{thesis}. To put the result into context, we estimated the corresponding normalization factor if more energy should go to the accelerated proton flux for the latter model. The experimental upper limit constrains the prediction made in~\cite{fargion} when assuming an average neutrino energy of 140~MeV. The still optimistic 500~MeV line is slightly below the reach of the current sensitivity. The second prediction~\cite{thesis}, however, is far below the current reach of IceCube, even when assuming the entire energy released during the solar flare goes to protons that subsequently produce pions. Similar results have been obtained for the other solar flares in Table~\ref{table:timewindow} and are discussed in~\cite{icrc-sf,thesis}.
\begin{table*}[t]
\begin{center}
\caption{Number of off-source and on-source IceCube events as well as the corresponding significance and flux upper limit obtained for each solar flare. For comparison, we show the expected number of events from the null hypothesis of no signal (Expected $N_{\text{on}}$) that depends on the data rate and the considered duration for each solar flare.}
\begin{tabular}{c| c | c | c| c| c|c|c}\hline
{Event} &$t_{\text{on}}$& $N_{\text{off}}$ & {{$N_{\text{on}}$} } &{{Expected}} &{Significance} &  Spectral &{Upper limit}  \\

&{(min)}&  &  &{{$N_{\text{on}}$} } &{$S$} & index $\delta_{\nu}$& 90\% C.L.(m$^{-2}$)  \\

\hline
&&&&&&4& 26 \\
Mar 7th, 2012 & 40&761& 67 & 62 &0.43 $\sigma$ && \\
&&&&&&6& 38\\
\hline
&&&&&&4& 23 \\
Feb 25th, 2014 &25 &611 & 27& 32 & 0.86 $\sigma$ && \\
&&&&&&6& 33 \\
\hline
&&&&&&4& 16\\
Sep 1st, 2014 & 14& 621 &21 & 18 & 0.65 $\sigma$&& \\
&&&&&&6& 23 \\
\hline
&&&&&&4& 131 \\
Sep 6th, 2017 & 517&  569& 639 & 620&  0.79 $\sigma$ &&\\
&&&&&&6& 192 \\
\hline
&&&&&&4&10 \\
Sep 10th, 2017 & 5.96&  529& 5 & 6 &0.64 $\sigma$ &&\\
&&&&&&6& 14 \\
\hline
\end{tabular}
\label{tab:resultsnumberofevents}
\end{center}
 \end{table*}
 
\section{Outlook}

In this paper, we presented the first results from a search for neutrinos coincident with solar flares using IceCube.  This was possible with the development of a new low-energy selection that allows IceCube to be sensitive to $<$5~GeV neutrinos.
We note that similar analyses may be carried out on other types of transient sources, such as compact binary mergers, fast-radio bursts, or novae. \\
The limitation of this analysis is given by the current effective area of the IceCube detector for GeV neutrinos. Besides the small cross section of GeV neutrinos with the surrounding matter that limits the number of interactions, each of these interactions needs to trigger the detector to be recorded. 
As previously described, the minimal requirement to be considered as an event is 3 DOMs with HLC pulses within 2.5~$\mu$s. 

A future option may be to utilize the HitSpooling data, which saves every single hit occurring in the detector, independent of trigger condition~\cite{jinst}. This means that sub-threshold neutrino interactions, lost in regular IceCube data, can be saved and studied. Using such a data stream would directly result in an increase of the sensitivity. HitSpool data structures have a significantly larger size than regular IceCube data, and cannot therefore be continuously saved. To take advantage of this new data stream and the increased sensitivity it offers, we have created an alert system based on  Fermi-LAT data. The system is continuously searching for significant solar flare events in Fermi-LAT data in view of triggering the IceCube HitSpool data stream~\cite{jinst}.  This system is running since September 2015 and successfully saved data for the bright solar flares of September 2017. A dedicated analysis of solar flares using HitSpooling is currently ongoing.

The landscape of large neutrino telescopes is expected to widen in the coming years, with among others, the deployment of the IceCube Upgrade~\cite{icrc-upgrade} within IceCube. This detector will demonstrate a lower detection threshold together with enhanced reconstruction capabilities because of the multi-photomultiplier geometry of its sensors. We are exploring how the analysis of the upcoming $25^{th}$ solar cycle would benefit from the dense instrumentation and the new sensors.

\begin{acknowledgements} 
The IceCube collaboration acknowledges the significant contributions to this manuscript from Gwenha\"el de Wasseige. We thank Melissa Pesce-Rollins for fruitful discussion. We gratefully acknowledge the support from the following agencies and institutes: USA {\textendash} U.S. National Science Foundation-Office of Polar Programs, U.S. National Science Foundation-Physics Division, Wisconsin Alumni Research Foundation, Center for High Throughput Computing (CHTC) at the University of Wisconsin{\textendash}Madison, Open Science Grid (OSG), Extreme Science and Engineering Discovery Environment (XSEDE), Frontera computing project at the Texas Advanced Computing Center, U.S. Department of Energy-National Energy Research Scientific Computing Center, Particle astrophysics research computing center at the University of Maryland, Institute for Cyber-Enabled Research at Michigan State University, and Astroparticle physics computational facility at Marquette University; Belgium {\textendash} Funds for Scientific Research (FRS-FNRS and FWO), FWO Odysseus and Big Science programmes, and Belgian Federal Science Policy Office (Belspo); Germany {\textendash} Bundesministerium f{\"u}r Bildung und Forschung (BMBF), Deutsche Forschungsgemeinschaft (DFG), Helmholtz Alliance for Astroparticle Physics (HAP), Initiative and Networking Fund of the Helmholtz Association, Deutsches Elektronen Synchrotron (DESY), and High Performance Computing cluster of the RWTH Aachen; Sweden {\textendash} Swedish Research Council, Swedish Polar Research Secretariat, Swedish National Infrastructure for Computing (SNIC), and Knut and Alice Wallenberg Foundation; Australia {\textendash} Australian Research Council; Canada {\textendash} Natural Sciences and Engineering Research Council of Canada, Calcul Qu{\'e}bec, Compute Ontario, Canada Foundation for Innovation, WestGrid, and Compute Canada; Denmark {\textendash} Villum Fonden and Carlsberg Foundation; New Zealand {\textendash} Marsden Fund; Japan {\textendash} Japan Society for Promotion of Science (JSPS) and Institute for Global Prominent Research (IGPR) of Chiba University; Korea {\textendash} National Research Foundation of Korea (NRF); Switzerland {\textendash} Swiss National Science Foundation (SNSF); United Kingdom {\textendash} Department of Physics, University of Oxford. G. de Wasseige acknowledges support from the European Union's Horizon 2020 research and innovation programme under the Marie Sklodowska-Curie grant agreement No 844138.\end{acknowledgements}

\end{document}